# Twitter Spam and False Accounts Prevalence, Detection and Characterization: A Survey

*Emilio Ferrara*
*Annenberg School for Communication and Journalism, University of Southern California*
*Dept. of Computer Science, Viterbi School of Engineering, University of Southern California*
*Information Sciences Institute, University of Southern California*

## Abstract

The issue of quantifying and characterizing various forms of social media manipulation and abuse has been at the forefront of the computational social science research community for over a decade. In this paper, I provide a (non-comprehensive) survey of research efforts aimed at estimating the prevalence of spam and false accounts on Twitter, as well as characterizing their use, activity, and behavior. I propose a taxonomy of spam and false accounts, enumerating known techniques used to create and detect them. Then, I summarize studies estimating the prevalence of spam and false accounts on Twitter. Finally, I report on research that illustrates how spam and false accounts are used for scams and frauds, stock market manipulation, political disinformation and deception, conspiracy amplification, coordinated influence, public health misinformation campaigns, radical propaganda and recruitment, and more. I will conclude with a set of recommendations aimed at charting the path forward to combat these problems.

## Introduction

Hailing as harbingers of democracy during the late 2000s [Howard et al., 2011], social media have increasingly demonstrated the potential to be abused, which can lead to harmful consequences for their users and society at large [Ferrara, 2015]. Many platforms, from fringe to mainstream, have demonstrably been the subject of some form of abuse or manipulation, as reported by practitioners, the research community, or the press, leding to potentially nefarious implications [Ferrara, 2019]: For example, mainstream platforms like Facebook, Youtube, or Reddit have been associated with the rampant spread of spam, radical content, misinformation, and conspiracies [Del Vicario et al., 2016; Tufekci, 2018; De Zeeuw et al., 2020]. Spam was particularly pronounced during the COVID-19 pandemic, and multimedia-based platforms such as Tiktok and Instagram became major drivers of public health misinformation [Shahsavari et al., 2020; Basch et al., 2021; Quinn et al., 2021]. Numerous studies attributed the incubation of malicious information to fringe communities: from niche platforms and discussion forums such as Gab, Minds, Parler, Voat, and 4Chan, to the alt-right information ecosystem of podcasts and fake news outlets like Infowars and Breibart, misinformation and conspiracies born out of small radical groups systematically made it to the mainstream [Zannettou et al., IMC 2018; Zannettou et al., WWW'18; Tangherlini et al., 2020; Papasavva et al., 2021; Aliapoulios et al., 2021].

But chief among all platforms, for the extent of reported spam, manipulation and abuse, might be Twitter: possibly in part due to the ease of access to its data [Williams et al., 2013], and in part because it has been under public scrutiny via multiple government and federal investigations, Twitter attracted an overwhelming amount of attention by researchers. The platform has been the subject of thousands of academic studies aimed at highlighting a multifaceted typology of problematic behaviors [Zimmer & Proferes, 2014].

It would impossible to summarize this whole literature in one single review: for this reason, in this survey, I will focus on the Twitter-related literature with a special focus on the problem of spam and false accounts, their behavior characterization, detection, use and effects. This choice is dictated by the ample amount of research that investigates the use of spam, false and compromised accounts, as well state-sponsored troll farms, bot and automated accounts, in the context of nefarious activities like scams, financial frauds, coordinated political campaigns, orchestrated misinformation, crises response disruption, radical propaganda, etc.

## Background

### A.     Taxonomy of Spam and False Accounts

Twitter defines four types of behavior as spam in their policy: (a) "commercially-motivated spam, that typically aims to drive traffic or attention from a conversation on Twitter to accounts, websites, products, services, or initiatives;" (b) "inauthentic engagements, that attempt to make accounts or content appear more popular or active than they are;" (c) "coordinated activity, that attempts to artificially influence conversations through the use of multiple accounts, fake accounts, automation and/or scripting;" and, (d) "coordinated harmful activity that encourages or promotes behavior which violates the Twitter Rules."

Such violations can be enacted by multiple sources, including but not limited to (i) promotional accounts (spammers); (ii) false/fake accounts, (iii) bots, or automated accounts; (iv) accounts with malicious intent (sometimes referred to as trolls); (v) compromised and repurposed accounts.

I provide a taxonomy of such accounts next, with the aim to later map back to this taxonomy when discussing the capability of state of the art machine learning models, and the findings of research studies in regard to the extent of different forms of spam on Twitter:

- **Promotional accounts** (or **spammers**): The category encompasses any account that is created on the platform with the deliberate intention to spam, i.e., promote commercial products or services, with the goal of driving other (legitimate) users' attention to that. This type of account can be controlled by human users, can be automated, or a hybrid of both. They tend to be disproportionately more active than most typical legitimate Twitter users.

- **False (**or **fake) accounts**: The category comprises any account that is created on the platform with the intention of boasting the visibility or appearance of influence of other accounts. False accounts are often sold as fake followers, i.e., pay-per-follower services exist that allow Twitter users to purchase followers that will artificially inflate their apparent visibility and popularity. Fake accounts are often inactive or minimally active, they engage predominantly in following other users, and occasionally on rebroadcasting (retweeting) such users' posts, or (sometimes automatically) liking the content produced by the users they are paid to follow.

- **Bots** (or **automated**) **accounts**: This category distinguishes user accounts based on the way their account is controlled. A bot account is an account that is exclusively or predominantly controlled by software, i.e., more or less complex computer scripts that automate the activity and behavior of the account with the goal to mimic, more or less faithfully, the way a human user would behave on the platform. Some bots are disproportionately more active than most human users, since their activity is automated.

Simple bots have limited functionalities such as automatically retweeting or liking certain content, whereas bots that rely on more advanced Artificial Intelligence, are capable of creating content that is often indistinguishable from human-generated content. Bots can also mimic or "clone" the behavior of target human users by simply imitating their temporal and network activities to simulate the patterns of engagement of real human users on Twitter.[1]

● **Malicious** (or **troll**) **accounts**: This category distinguishes users accounts based on their intent. A common practice in the literature is to consider malicious users to be engaged in inauthentic activities when they are purposely initiating or following an agenda, i.e., political propaganda, conspiratorial campaigns, disinformation, influence operations, toxic attitudes, etc., aimed to artificially steer or distort a Twitter conversation, and/or to deceive legitimate users therein involved. Malicious accounts are often engaged in campaigns that can be organized by state-sponsored foreign governments, influence groups, and occasionally by lone wolves. They are also often engaged in coordinated activity carried out in orchestration with many other such accounts. Malicious accounts can be automated, but are often controlled by human operators, sometimes referred to as trolls (e.g., Russian Internet Research Agency troll factory).

● **Compromised accounts**: This category comprises accounts that were originally created by legitimate users but have since been hacked or compromised, and their control has been passed to somebody else. The original owner might or may not have access to such accounts, hence might be unwillingly and unknowingly complicit in the misuse of their accounts. Often these accounts are dormant and the owner may not notice. Compromised accounts are often sold on black markets, e.g., the Dark Web. They are also often repurposed into botnets.

## B. Creation Of Spam Accounts

To understand how spam accounts behave, and how to detect and curb them, it is useful to first understand how such accounts are created and operated in the first place.

Of the taxonomy above, special attention should be paid to bot accounts, which operate in a manner substantially different from the other four categories of spam accounts, i.e., by means of software automation. The other categories of accounts are usually created and operated manually (i.e., by human users) and their detection warrants a separate discussion, so I will focus on bots.

---

[1] The definition of what a social bot is has seen some debate in the literature. A bot (short for robot) generally refers to an entity operating in a digital space that is controlled by software rather than human. Bots have been categorized according to various taxonomies [R. Gorwa and D. Guilbeault; Policy & Internet 2018], [S. Woolley; First Monday 2016]. In this taxonomy, I use the term bot as a shorthand to social bot, a concept that refers to a social media account controlled, predominantly or completely, by computer software (a more or less sophisticated artificial intelligence), in contrast with accounts controlled by human users. I introduced this definition in [E. Ferrara et al., CACM 2016], which to date remains by far the most cited article on the subject of Twitter bots, with over 2,000 citations. This definition is in line with the recommendations of [R. Gorwa and D. Guilbeault; Policy & Internet 2018] who provided the most comprehensive survey on the typologies of bots (cf., page 9: "*we suggest that automated social media accounts be called social bots*").

## 1.     How to Create a Bot

Early social media bots, in the late 2000s, were created to tackle simple tasks, such as automatically retweeting content posted by a set of sources or finding and posting news from the Web. Today, the capabilities of bots have significantly improved: bots rely on the fast-paced advancements of Artificial Intelligence, especially in the area of natural language generation, and use pre-trained multilingual models like OpenAI's GPT-2 [A. Radford et al., 2019] and its evolutions or alternatives, to generate human-like content. This framework allows the creation of bots that generate genuine-looking short texts on platforms like Twitter, requiring revised strategies to distinguish between human and automated accounts [Alarifi et al., Inf. Sci. 2016].

The barriers to bot creation and deployment, as well as the required resources to create large bot networks, have also significantly decreased: for example, it is now possible to rely upon bot-as-a-service (BaaS), to create and distribute large-scale bot networks using pre-existing capabilities provided by companies like ChatBots.io, and run them in cloud infrastructures like Amazon Web Services, Heroku, or behind proxy services, e.g., residential IP proxies (RESIP) such as Brigth Data, SmartProxy, ProxyRack, or OxyLabs [Chiapponi et al., 2022] in an attempt to disguise their actions.

## 2.     Open Source Twitter Bots

A recent survey discusses readily-available Twitter bot-making tools [F. Daniel & A. Millimaggi; J. Web Eng. 2020]: The authors provide an extensive overview of open-source GitHub repositories and describe how prevalent different automation capabilities, such as tweeting or resharing, are across these tools.

According to Daniel and Millimaggi, whose survey focused exclusively on repositories for Twitter bots developed in Python, there are hundreds of such open-source tools readily available for deployment. The authors studied 60 such bot-making tools and enumerated the most common capabilities. Typical Twitter bots' automated features include:

- Searching users, trends, and keywords;

- Following users, trends, and keywords;

- Liking content, based on users, trends, and keywords;

- Tweeting and mentioning users and keywords, based on AI-generated content, fixed-templated content, or cloned-content from other users;

- Retweeting users and trending content, and mass tweeting;

- Talking to (replying) other users, based on AI-generated content, templated content, or cloned-content from other users; finally,

- Pausing activity to mimic human circadian cycles and bursty behaviors, as well as to satisfy API constraints, and to avoid suspension.

According to [F. Daniel & A. Millimaggi; J. Web Eng. 2020], these features can enable bots to carry out various forms of abuse including: denigrate and smear, deceive and make false allegations, spread misinformation and spam, and finally clone users and mimic human interests.

## C.    Detection of Spam Accounts

Most of the work to develop models to automatically detect spam accounts has focused on the detection of bots. This is in part due to the fact that the definition of spam oftentimes entails a normative judgment of the intent behind an account's activity and behavior (i.e., inferring whether the entity behind the account is attempting to gain from the spamming activity). Conversely, the definition of bot can be operationally defined unequivocally as 'software-operated' or 'automated' account (as opposed to human-operated account) – irrespective of the account's intent (hence, bots are not to be considered inherently 'bad' or ill-intended, although research shows that the vast majority of bot accounts attempt to disguise or hide their automated nature [Ferrara et al., 2016]).

### 1.    How to Detect Bots

The problem detection and suspension of bots on social media platforms like Twitter has been well studied for over a decade [K. Thomas et al., SIGCOMM 2011; Pierri et al., 2022], [Y. Boshmaf et al., Comput. Networks 2013], [C. Freitas et al., ASONAM 2015]. Some popular bot detection techniques based on machine learning have been pioneered by groups at Indiana University, University of Southern California, and University of Maryland, in the context of the SMISC (Social Media in Strategic Communication) program sponsored by DARPA (the U.S. Defense Advanced Research Projects Agency):[2] In 2015, the DARPA SMISC program organized a DARPA Challenge aimed at detecting bots used within anti-vaccination campaigns on Twitter [V. S. Subrahmanian et al., Computer 2016]. More traditional cybersecurity techniques led to the discovery of large bot networks (botnets) on Twitter by various research groups [Y. Boshmaf et al., ACSAC 2011], [N. Abokhodair et al., CSCW 2015], [J. Echeverria & S. Zhou; 2017].

### 2.    Tools for Bot Detection

The literature on bot detection is extensive. In [E. Ferrara et al., CACM 2016], we proposed a simple taxonomy to divide bot detection approaches into three classes: (1) systems based on social network information; (2) systems based on crowd-sourcing and the leveraging of human intelligence; (3) machine learning methods based on the identification of highly-predictive features that discriminate between bots and humans. Undoubtedly, other taxonomies that propose alternative or complementary criteria also exist [R. Gorwa and D. Guilbeault; Policy & Internet 2018], [S. Woolley; First Monday 2016].

Some openly accessible tools exist to detect bots on platforms like Twitter:

- Botometer,[3]  a  bot-detection  tool  developed  at  Indiana  University (https://botometer.osome.iu.edu/faq);

---

[2]  DARPA  Social  Media  in  Strategic  Communication  (SMISC)  program  (archived)  website: https://www.darpa.mil/program/social-media-in-strategic-communication
[3] Description of Botometer taken from the official website: "Botometer is a machine learning algorithm trained to calculate a score where low scores indicate likely human accounts and high scores indicate likely bot accounts. To calculate the score, Botometer compares an account to tens of thousands of labeled examples. When you check an account, your browser fetches its public profile and hundreds of its public tweets and mentions using the Twitter API. This data is passed to the Botometer API, which extracts over a thousand features to characterize the

- BotometerLite,[4] a lightweight bot-detection model developed by the same team of Botometer (https://botometer.osome.iu.edu/botometerlite);

- Bot Sentinel,[5] a public dataset of accounts classified as bots or not via machine learning (https://botsentinel.com);

- Tweet Bot or Not,[6] an R package for classifying Twitter accounts as bot or not, (https://github.com/mkearney/tweetbotornot);

- DeBot,[7] a real-time bot detection system by Univeristy of New Mexico's researchers (https://www.cs.unm.edu/~chavoshi/debot).

Various models have been proposed to detect bots using sophisticated machine learning techniques, such as:

- Deep learning [S. Kudugunta and E. Ferrara; Inf. Sci. 2018];

- Anomaly detection [A. Minnich et al., ASONAM 2017], [Gilani et al., ASONAM 2017], [J Echeverria et al., ACSAC 2018].

- Graph-based techniques [Feng et al., CIKM 2021], [Feng et al., 2022];

- Time series analysis [N. Chavoshi et al., ICDM 2016], [D. Stukal et al., Big Data 2017], [I Pozzana & E Ferrara; Frontiers in Physics 2020].

Finally, numerous recent studies have used mixed approaches that combine some of the ideas above or build upon state-of-the-art advances in artificial intelligence techniques. This includes (i) use of long-short term memory (LSTM) neural networks [Wei et al., 2019]; (ii) generative

---

account's profile, friends, social network structure, temporal activity patterns, language, and sentiment. Finally, the features are used by various machine learning models to compute the bot scores. We do not retain any data other than the account's ID, scores, and any feedback optionally provided by the user."

[4] Description of BotometerLite taken from the website: "BotometerLite is a lightweight and scalable bot detector for Twitter. Unlike Botometer, BotometerLite only requires one tweet to perform bot detection, which gives it the following advantages: 1. It can efficiently analyze large amounts of accounts; 2. It allows bot detection on historical data, whereas Botometer only returns the most recent status."

[5] Description of Botsentinel taken from the official Website: "Bot Sentinel is a free non-partisan platform developed to classify and track inauthentic accounts and toxic trolls. The platform uses machine learning and artificial intelligence to classify Twitter accounts, and then adds the accounts to a publicly available database that anyone can browse."

[6] Description of TweetBotOrNot taken from the official website: "Uses machine learning to classify Twitter accounts as bots or not bots. The default model is 93.53% accurate when classifying bots and 95.32% accurate when classifying non-bots. The fast model is 91.78% accurate when classifying bots and 92.61% accurate when classifying non-bots."

[7] Description of DeBot taken from official website: "DeBot is real-time bot detection system. The project started on Feb 2015 and it has been collecting data since Aug 2015. High correlation in activities among users in social media is unusual and can be used as an indicator of bot behavior. DeBot identifies such bots in Twitter network. Our system reports and archives thousands of bot accounts every day. DeBot is an unsupervised method capable of detecting bots in a parameter-free fashion. In March 2017, DeBot has collected over 730K unique bots. Since we are detecting and archiving Twitter bots on a daily basis, we can offer two different service: bot archive API and on-demand bot detection platform."

adversarial networks (GANs) [Wu et al., 2020]; (iii) computational stylometrics [Cardaioli et al., 2021]; (iv) random walk based detection [Fukuda et al., 2022]; (v) embedding based techniques [Shubham et al., 2021]; (vi) transformer based techniques [Martin-Gutierrez et al., 2021]; (vii) learning automata [Rout et al., 2020]; (viii) attention mechanisms [Shen et al., 2022]; (ix) ensembles models [Hrushikesh et al., 2021].

## 3.    Botometer

The first tool to become publicly available for computational bot detection is Botometer, originally released as "BotOrNot" back in 2014. This is the tool that has been used to complement bot analysis in this study, and it provides some of the state of the art techniques in computational bot detection. Botometer has been openly available to the public since its inception release in mid 2014: The tool's Version-1 was accompanied by a peer reviewed paper that describes the methodology [Davis et al., WWW 2016].

Botometer continually undergoes through improvements and the Indiana University team in charge of this tool periodically releases new and improved versions to the public. **Figure 1** illustrates the timeline of Botometer versions published since 2014 to date.

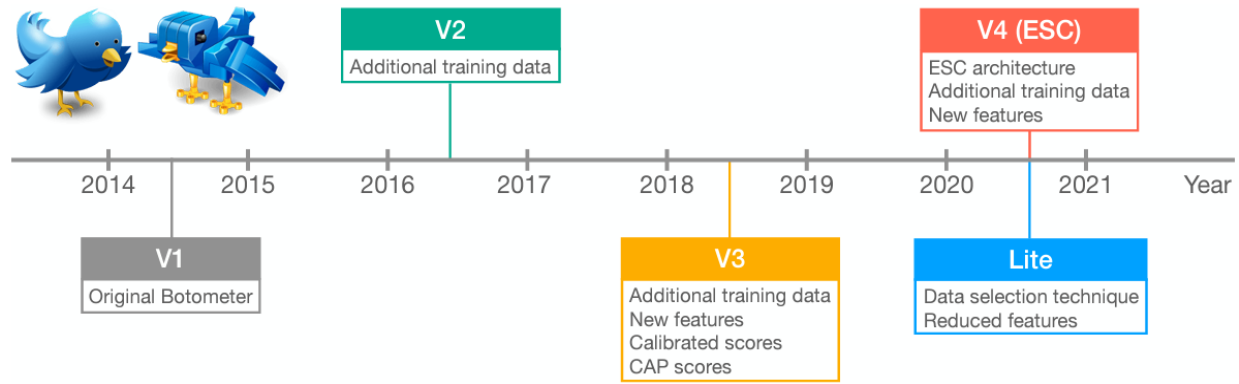

**Figure 1**: The timeline of Botometer (formerly BotOrNot) versions. From [Yang et al., JCSS 2022].

Botometer's authors define it as "a machine learning tool for bot detection on Twitter" [Yang et al., JCSS 2022]. The current version of Botometer is Version 4 (v4 ESC, *i.e., Ensemble of Specialized Classifiers*). Adopting Botometer can have several advantages, as portrayed by the same paper (verbatim) that describes the tool:

*"First, it is well maintained and has been serving the community for the past seven years without major outages. It has also been routinely upgraded to stay accurate and relevant. Second, Botometer is easily accessible through both a web interface and an application programming interface (API). Anyone with a Twitter account can use the web version for free; researchers with Twitter developer accounts can use the API endpoints to analyze large-scale datasets. The API has a nominal fee for heavy use, which discourages abuse and partially offsets infrastructure and maintenance costs. Third, Botometer is quite popular. It handles around a quarter million daily queries—over half a billion in total since its inception. Finally, Botometer has been extensively validated in the field. Many researchers have applied Botometer in their studies to directly investigate social bots and their impact [Keller & Klinger, 2019; Broniatowski et al., 2018; Allem et al., 2020; Fan et al., 2020], or to distinguish human accounts and bot-like accounts in order to*

*better address their questions of interest [Vosoughi et al., 2018; Grinberg et al., 2019; Bovet & Makse, 2019]."*

## 4.    How Botometer Works

The latest version of Botometer (V4 ESC) is described as follows in the related peer reviewed paper accompanying the tool [Yang et al., JCSS 2022]:

- *"Under the hood, Botometer is a supervised machine learning classifier that distinguishes bot-like and human-like accounts based on their features (i.e., characteristics). Unsupervised methods have also been proposed in the literature [Chavoshi et al., 2016; Echeverria & Zhou, 2017], but they only allow for the detection of specific, predefined behaviors. Therefore they are not suitable to build a general detection tool."*
- *"Botometer considers over 1,000 features that can be categorized into six classes: user profile, friends, network, temporal, content and language, and sentiment [Varol et al., 2017]. For example, the user profile category includes features such as the length of the screen name, whether the account uses the default profile picture and background, the age of the account, etc. The content and language category consists of features such as the number of verbs, nouns, and adjectives in the tweets. For a given account, these features are extracted and encoded as numbers. This way the account can be represented by a vector of feature numbers, enabling machine learning classifiers to process the information."*

## 5.    Accuracy and Robustness of Bot Detection Tools

The performance and accuracy of different bot detection models have been tested extensively by the author of such models, whose details appear published in peer reviewed academic papers, as well as independently verified by other researchers besides the original developers in further studies [J. Echeverria et al., ACSAC 2018], [Alipur et al., ARES 2022].

### a)    LOBO - A System to Evaluate the Generalizability of Bot Detection Tools

In 2018, [J. Echeverria et al., ACSAC 2018] published a first independent and rigorous evaluation of the accuracy of Botometer. This test was based on Version-2 of Botometer (to date, Botometer reached Version 4, see [Yang et al., JCSS 2022]). They used a plethora of datasets and variations of the bot detection task at end, and reached the following conclusions: (verbatim)

> *"This can be seen in **Table 1**, which shows the average botometer scores for each of the bot classes. To evaluate whether this tool would be able to predict the dataset, we provide another metric which is the percentage of the queried accounts that receive a botometer score over 0.5 . Because of rate limiting, we only collected botometer scores for up to 1,000 randomly selected accounts belonging to each of the classes. We can see that many of these bot classes are overwhelmingly classified as users, for example, only 2.75% of the Bursty Bots are classified as bots, and less than 10% of DeBot bots (dataset C) are classified as bots. Both of them with average botometer scores less than 0.1, indicating that they are "very likely" to be users. Different bot classes will achieve reasonable and even perfect performance on this bot detection task, but variability between bot classes is very clear."*

| ID | Name | BTS(%) | BTS(Avg) | Size |
|----|------|--------|----------|------|
| A | Star Wars Bots | — | — | 357,000 |
| B | Bursty Bots | 2.75 | 0.04 | 500,000 |
| C | DeBot | 7.67 | 0.09 | 700,000 |
| D | Fake followers | 96.79 | 0.90 | 721 |
| E | Social spambots #1 | 92.35 | 0.85 | 551 |
| F | Social spambots #2 | 99.37 | 0.96 | 3,320 |
| G | Social spambots #3 | 94.10 | 0.87 | 458 |
| H | Traditional spambots #1 | 98.28 | 0.93 | 872 |
| I | Traditional spambots #2 | 100.00 | 0.85 | 1 |
| J | Traditional spambots #3 | 66.08 | 0.60 | 283 |
| K | Traditional spambots #4 | 97.81 | 0.90 | 977 |
| L | ~ 1k followers | 20.89 | 0.21 | 387 |
| M | ~ 100K followers | 10.90 | 0.13 | 534 |
| N | ~ 1M followers | 1.32 | 0.02 | 229 |
| O | ~ 10M followers | 0.00 | 0.00 | 26 |
| Q | Fake followers-FSF | 100.00 | 0.96 | 33 |
| S | Fake followers-INT | 100.00 | 0.95 | 64 |
| T | Fake followers-TWT | 95.34 | 0.89 | 624 |
| V | HoneyPot bots (Darpa) | 27.69 | 0.30 | 2,521 |
| W | Attack on Ben Nimmo | 59.09 | 0.54 | 1,558 |
| X | Attack on Brian Krebs | 83.05 | 0.78 | 728 |

**Table 1:** Performance of Botometer (v2) on different datasets; BTS(%) is the proportion of correctly detected bots, BTS(Avg) is the average bot score. From [Echeverria et al., 2018].

The general performance of Botometer in this independent analysis is very high, reaching on average over 90% accuracy on most bot datasets. Nevertheless, this study highlighted one limitation of the original versions of Botometer, which was the lower performance in generalizing to unseen bot types (e.g., the DeBot dataset). This particular problem was addressed by Botometer V3, using the technique illustrated in the paper titled *'Scalable and generalizable social bot detection through data selection'* [Yang et al., AAAI 2020].

**b)      State-of-the-art Bot Detection Benchmark**

In 2022, [Alipur et al., ARES 2022], set to independently test the accuracy of two state-of-the-art (SOTA) Bot Detection tools, Botometer and TweetBotOrNot. They used over a dozen datasets, including the DeBot data [J. Echeverria et al., ACSAC 2018], and tested various generalizability and robustness aspects of bot detection. Their analysis reached the following conclusions.

**6.      Botometer Results**

Here, two excerpts, verbatim from their analysis and discussion of Botometer performance:

- "As shown in **Table 2A** and **Table 2B** the Botometer performs well in identifying bot and human accounts on the data on which it is trained, especially for the threshold of 0.5."

(A)

| Dataset | >=0.5 | >=0.75 | >=0.8 |
|---|---|---|---|
| TwiBot-20 | 30% | 18% | 16% |
| Cresci-rtbust-2019 | 80% | 70% | 64% |
| Botometer-feedback-2019 | 82% | 64% | 59% |
| Kaiser | 87% | 79% | 76% |
| Vendor-purchased-2019 | 80% | 41% | 33% |
| Pronbots-2019 | 90% | 84% | 82% |
| Astroturf | 74% | 63% | 58% |
| Political-bots-2019 | 100% | 46% | 38% |
| Botwiki-2019 | 99% | 97% | 96% |

(B)

| Dataset | <0.5 | <0.75 | <0.8 |
|---|---|---|---|
| TwiBot-20 | 78% | 89% | 91% |
| Cresci-rtbust-2019 | 65% | 80% | 86% |
| Botometer-feedback-2019 | 81% | 90% | 92% |
| Kaiser | 75% | 92% | 94% |
| Celebrity-2019 | 94% | 98% | 98% |
| Verified-2019 | 91% | 98% | 99% |

**Table 2**: Percentage of (A) bots and (B) humans detected by Botometer according to training dataset and threshold. From [Alipur et al., ARES 2022].

- "In addition to the datasets that the Botometer was trained on, we used other datasets that Juan Echeverria et al. used in their research in 2018 [J. Echeverria et al., ACSAC 2018]. We call these datasets Juan datasets. As illustrated in **Table 2**, Botometer still performed well except on the DeBot dataset. DeBot dataset was collected by Juan Echeverria et al. by querying the accounts detected as bots by the DeBot Bot Detection Service. It should be noted here that Botometer performed better on this dataset in our replication study than the results reported by Encheverra et al. when they performed the original study [J. Echeverria et al., ACSAC 2018]. on the same datasets. This might be because of the improvements of the Botometer's detection module."

## 7. TweetBotornot Results

Here, two excerpts, verbatim from their analysis and discussion of TweetBotOrNot performance:

- "**Table 3A** and **Table 3B** depict Tweetbotornot's performance on both Botometer trained and Juan datasets, respectively. These results show that Tweetbotornot does not perform as well as Botometer on these datasets."

(A)

| Dataset | >0.5 | >0.6 | >0.7 | >0.8 |
|---|---|---|---|---|
| TwiBot-20 | 4% | 4% | 3% | 3% |
| Cresci-rtbust-2019 | 2% | 2% | 2% | 2% |
| Botometer-feedback-2019 | 30% | 30% | 29% | 29% |
| Kaiser | 69% | 68% | 68% | 67% |
| Vendor-purchased-2019 | 4% | 3% | 3% | 3% |
| Pronbots-2019 | 2% | 2% | 2% | 1% |
| Astroturf | 1% | 1% | 1% | 1% |
| Political-bots-2019 | 31% | 31% | 31% | 23% |
| Botwiki-2019 | 99% | 99% | 99% | 99% |

(B)

| Dataset | >0.5 | >0.6 | >0.7 | >0.8 |
|---|---|---|---|---|
| BurstyUsers | 2% | 2% | 2% | 2% |
| Journalist attack bots (Ben Nimmo) | 2% | 2% | 2% | 1% |
| Journalist attack bots (Brian) | 2% | 2% | 1% | 1% |
| DeBot | 4% | 4% | 4% | 3% |

**Table 3**: Percentage of bots detected by Botornot according to Botometer (A) and Juan (B) training dataset and threshold. From [Alipur et al.; ARES 2022].

- "*Among the datasets we examined, there is only one, namely Kaiser dataset, on which Botometer and Tweetbotornot detect a relatively similar number of accounts. This* *similarity in the performance led us to examine whether the bots detected by the Botometer are the same as the bots detected by the Tweetbotornot. Researchers in [Echeverria et al., 2018] showed that Tweetotornot and Botometer had very little overlap in the bots identified in political data, especially those that contained certain hashtags. However, in our study of different datasets regardless of the content of the tweets, Tweetbotornot and Botometer overlap a lot in the detected bots. In Table 8,[8] we show the number of records in each dataset, the number of bots correctly detected by Botometer at a threshold of >=0.5, the number of bots detected by Tweetbotornot with a bot-probability of >0.5, and the number of common bots (cb) that both systems detected based on the same threshold and probability. According to these results, most of the bots detected by both systems are the same. In general, there is a high overlap in the number of bots detected by Tweetbotornot with the ones Botometer detected. Most of the time Tweetbotornot detected bots are a subset of Botometer detect bots, column cb in Table 8."*

## 8. Overall Conclusions of the 2022 SOTA Bot Detection Benchmark

These are the overall conclusion of the 2022 SOTA Bot Detection Benchmark study [Alipur et al., ARES 2022], verbatim:

*"Our results show that the performances of SOTA systems in detecting bot accounts are better than detecting human accounts. Moreover, the performances of SOTA systems do have a substantial overlap in terms of the bots as well as the human accounts they are able to detect. However, the performances of SOTA systems decrease when they are evaluated on totally different datasets than the one used to train them. This decrease is bigger in the performance of Tweetbotornot compared to the performance of Botometer."*

## 9. On the Accuracy and Robustness of Computational Bot Detection Models

Based on the studies explained above, [J. Echeverria et al., ACSAC 2018] [Alipur et al., ARES 2022], it can be concluded that computational bot detection (i.e., bot detection based on machine learning models) is a valid scientific tool and that Botometer yields the most accurate classification among the assessed, publicly available tools. As such, Botometer represents the state of the art (SOTA) in computational bot detection to date. In most tests published by independent researchers, the accuracy of the tool is moderately to very high (above 80% for almost all data and tasks, with peaks above 95% for simpler classification tasks). Botometer is often the go-to tool for tasks that requires scalability (i.e., the processing of millions of accounts in limited time), albeit manual double-checking and fine-tuning of the models is always advised.

## D. Prevalence of Spam Accounts

## 1. Timeline of Recent Reports (Reverse Chronological Ordering)

Various studies have been published reporting on the number of spam accounts, typically focusing on promotional, fake accounts, and bots. Each study employs different methodologies, sample sizes, datasets, etc. In the following, we summarize some of the more transparent reports that appeared in the last 5 years.

---

[8] See Table 8 in the original paper [Alipur et al., ARES 2022], omitted here.

- In June 2022, we analyzed 5.5M users, contributing over 207M retweets and 23M replies, active in the context of the broad Covid19 discussion [Chang & Ferrara; 2022]. Here some highlights of the study:

  *"We randomly sampled 200 accounts (100 from each account) which were validated by three researchers. This yielded an average of 88% agreement with the Botometer on average. "*

  *"The distribution of Botscores within our dataset is visualized in **Figure 2**, which shows a natural bimodal break between 0.5. Hence, we label users with botscores above 0.5 as bots."*

  *"In absolute terms, liberal bots dominate the amount of retweets, taking up to 20% of all retweets. In contrast, conservative bots only amount to around 4.5% of the total human retweets."*

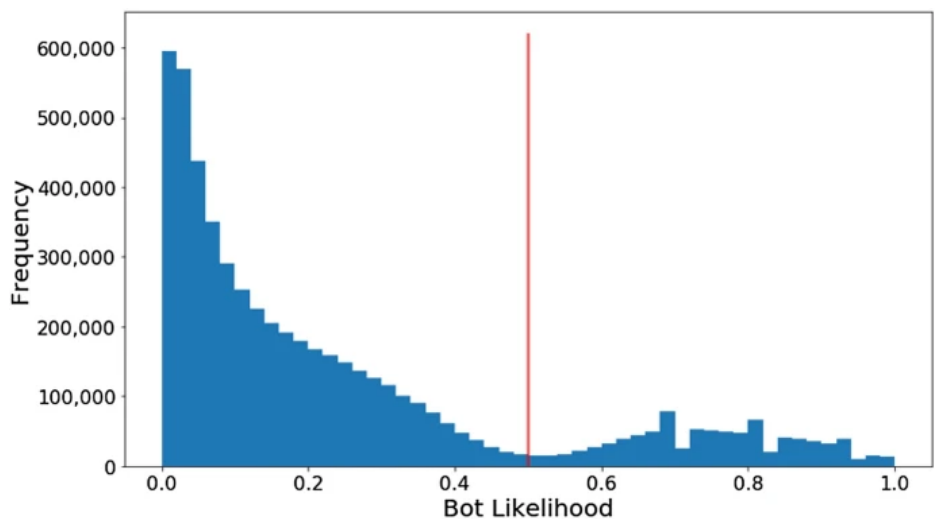

**Figure 2** shows the botscore distribution (which is the likelihood that a user is a bot), which forms a natural bimodal distribution break at 0.5. [Chang & Ferrara; JCSS

- In May 2019, Luceri and collaborators published a study of the prevalence of spam accounts in the context of the 2018 US midterms related discourse [Luceri et al., WWW 2019]. This is what they found (verbatim):

  *"we classified 21.1% of the accounts as bots, which in turn generated 30.6% of the tweets in our data set. Overall, Botometer did not return a score for 35,029 users that corresponds to 3.5% of the accounts"*

  *"99.4% of these accounts were suspended by Twitter, whereas the remaining percentage of users protected their tweets turning on the privacy settings of their accounts."*

- In November 2018, Stella and collaborators measured the amount of bots involved in the conversation about the 2017 Catalan referendum [Stella et al., PNAS 2018]; we reported that (verbatim):

> *"We monitored the Twitter stream and collected data by using the Twitter Search application programming interface (API), from September 22, 2017, to past the election day, on October 3, 2017 [...] This procedure yielded a large dataset containing ~3.6 million unique tweets, posted by 523,000 unique users."*

> *"Off-the-shelf learning models were trained on multiple historical ground truth datasets and achieved high detection accuracy (>90%) on cross-validation benchmarks. Logistic regression (LR), our reference model for this study, was selected for its best trade-off between scalability and accuracy: The model is very precise at detecting human accounts—precision rate (PR) 98%, compared with bot accounts (PR: 92%), while detecting nearly all existing bots—recall rate (RR) 99%, compared with human users retrieval (RR: 88%)."*

> *"We discovered that bots generated specific content with negative connotation that targeted the most influential individuals among the group of Independentists (i.e., Catalan independence supporters)."*

> *"bots produced 23.6% of the total number of posts during the event".*

- In May 2018, Pew Research reported on the study of the prevalence of bots-created tweeted links to news sites.[9] This is what they found (verbatim):

> *"Of all tweeted links to popular websites, 66% are shared by accounts with characteristics common among automated "bots," rather than human users."*

> *"Among popular news and current event websites, 66% of tweeted links are made by suspected bots – identical to the overall average. The share of bot-created tweeted links is even higher among certain kinds of news sites. For example, an estimated 89% of tweeted links to popular aggregation sites that compile stories from around the web are posted by bots."*

> *"A relatively small number of highly active bots are responsible for a significant share of links to prominent news and media sites. This analysis finds that the 500 most-active suspected bot accounts are responsible for 22% of the tweeted links to popular news and current events sites over the period in which this study was conducted. By comparison, the 500 most-active human users are responsible for a much smaller share (an estimated 6%) of tweeted links to these outlets".*

---

[9] Bots in the Twittersphere: An estimated two-thirds of tweeted links to popular websites are posted by automated accounts – not human beings https://www.pewresearch.org/internet/2018/04/09/bots-in-the-twittersphere

*"The study does not find evidence that automated accounts currently have a liberal or conservative "political bias" in their overall link-sharing behavior. This emerges from an analysis of the subset of news sites that contain politically oriented material. Suspected bots share roughly 41% of links to political sites shared primarily by liberals and 44% of links to political sites shared primarily by conservatives – a difference that is not statistically significant. By contrast, suspected bots share 57% to 66% of links from news and current events sites shared primarily by an ideologically mixed or centrist human audience."*

**Figure 3** illustrates the overwhelming prevalence of bot-produced links to popular news.

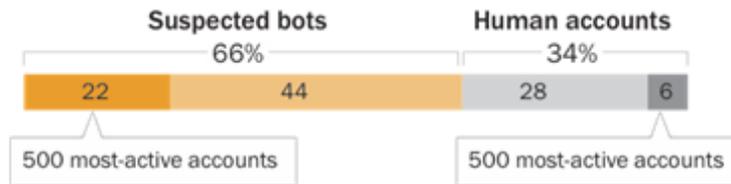

**Figure 3.**

Source: Analysis of 379,841 tweeted links to 925 popular news and current events websites collected over the time period July 27–Sept. 11, 2017.
"Bots in the Twittersphere"

**PEW RESEARCH CENTER**

- In August 2017, I published an analysis on disinformation and bot operations ahead of the 2017 French presidential election [Ferrara, First Monday 2017]. The study highlights that (verbatim):

*"By using a Logistic Regression model trained on the 10 metadata and activity features described above, we obtain very accurate user classification on the cross-validated tests (92 percent accuracy, 89 percent AUC-ROC). We adopted this model to detect all bots and separate them from human users in our dataset. "*

*"Out of 99,378 users involved in MacronLeaks, our model classified 18,324 of them as social bots, and the remainder of 81,054 as human users. The fraction of social bots amounts for about 18 percent of the total users involved in the campaign"*

*"Remarkably, among the top 15 social bots detected by our framework, four accounts have been so far deleted, seven have been suspended, and two have been quarantined by Twitter. Two accounts are still active, and they may be the result of misclassification. Overall, this example of manual verification suggests that we obtained 13 correct bots out of 15 detected, an accuracy of nearly 87 percent".*

- In May 2017, Varol and collaborators reported on the prevalence of bots on Twitter using a combination of machine learning bot detection and human validation [Varol et al., ICWSM 2017]. These are some of the findings (verbatim):

> *"We also estimated the fraction of bots in the active English-speaking population on Twitter. We classified nearly 14M accounts using our system and inferred the optimal threshold scores that separate human and bot accounts for several models with different mixes of simple and sophisticated bots. Training data have an important effect on classifier sensitivity. Our estimates for the bot population range between 9% and 15%."*

> *"Using mixtures with different ratios of accounts from the manually annotated and honeypot datasets, we obtain an accuracy ranging between 0.90 and 0.94 AUC."*

> *"Using an optimal threshold, we measured false positive and false negative rates at 0.15 and 0.11 respectively in our extended dataset. In these experiments, human annotation is considered as ground truth"*

- In November 2016, we published a study of social bot interference in the political discussion revolving around the 2016 US Presidential Election [Bessi & Ferrara; First Monday 2016]. In this study, we highlight that the fraction of bots engaged in the 2016 US presidential election was approximately 14.4% and they were responsible for approximately 18.45% of the content.

In **Table 4**, I summarize the findings of the studies above to draw a comprehensive conclusion about the estimated prevalence of spam or bot accounts on Twitter according to prior research:

| Study | Spam Account Prevalence | Spam Content Prevalence | Accuracy / Confidence | Focus |
|---|---|---|---|---|
| *[Chang & Ferrara; 2022]* | Unspecified | **24.5% of retweets** | 76% annotator agreement | Bots |
| *[Luceri et al., 2019]* | **21.1%** | **30.6%** | 99.4% | Bots |
| *[Stella et al., 2018]* | Unspecified | **23.6%** | 92-98% accuracy | Bots |
| *Pew Research 2018* | Unspecified | **66% of popular content** | 86% accuracy | Bots |
| *[Ferrara; 2017]* | **18.3%** | Unspecified | 87-92% | Bots/trolls |
| *[Varol et al., ICWSM 2017]* | **9-14%** | Unspecified | 90-94% accuracy; 88% annotator agreement | Bots/spammers |
| *[Bessi & Ferrara; 2016]* | **14.4%** | **18.45%** | 95% accuracy | Bots/trolls |

**Table 4**: Summary of the Studies on Spam and False Accounts Prevalence on Twitter.

The prevalence of accounts that engage in some type of spam activities, as defined by Twitter, is consistently reported between approximately 9 and 21 percent, with an average estimate at approximately 12.5% plus or minus 2% of the total Twitter accounts.

The prevalence of content that is considered spam has been measured on a broader range spanning from approximately 18% to 30% so it can be assumed to average at, or in excess of, 20% of the overall content appearing on Twitter.

The studies rely on a plethora of methodologies including but not limited to machine learning, human annotations, and combinations thereof, and when available, report accuracy and/or confidence typically above 80% and often above 90%, which suggests that the reported margin of error (both in favor and against spam classification, i.e., false positives and false negatives) renders the above consistent estimates reliable and robust.

## 2.　　Honeypots and Black Markets for Spam and False Accounts

There have been recent studies that attempted to identify spam accounts using strategies not based on machine learning or crowdsourcing. The two most common alternative strategies have included the use of honeypots to attract spammers, and the scrutiny of black markets (such as the dark Web) or services that provide spam and false accounts as a service for purchase. Here are some notable examples:

- Mazza and collaborators recently published the ReadyToAbuse study [Mazza et al., 2022], in which they tracked down tens of thousands of Twitter accounts (including compromised accounts) that are sold on the *buyaccs* website, a marketplace for purchasing fake and compromised accounts on various social media platforms, most prominently Twitter. This is what the study found (verbatim):

  *"We passively monitored a popular Twitter account merchant buyaccs.com from June 2019 to the end of July 2021 and detected 63,358 fake accounts for sale as a result"*

  *"We focused our analysis on the fake accounts that produced at least one tweet throughout 2020, identifying 5,457,758 tweets from 23,579 accounts;"*

  *"We uncovered four coordinated campaigns conducted by fake accounts and showed more details on their inauthentic behavior on Twitter"*

- In 2017, I published a study that unveiled the MacronLeaks disinformation campaign in the run-up to the 2017 French presidential election, and characterized the role that spam bots played in it. In the paper, besides unveiling the presence of over 18,000 bot accounts, I also documented the repurposing of at least 800 such accounts, which I found also active in the discourse surrounding the 2016 US Presidential Election. Investigative reports demonstrated that such accounts had been sold and purchased on the black markets, specifically on the Dark Web [Ferrara, First Monday 2017].

- In 2011, Lee and collaborators built a honeypot system designed to attract spammers on Twitter [Lee et al., AAAI 2011]. They found that (verbatim):

> *"The system ran from December 30, 2009 to August 2, 2010. During that time the social honeypots tempted 36,043 Twitter users, 5,773 (24%) of which followed more than one honeypot."*

> *"We found that Twitter eventually suspended the accounts of 5,562 (or 23% of the total polluters identified by the social honeypots). We observe that of the 5,562, the average time between the honeypot tempting the polluter and the account being suspended was 18 days."*

## E.    Effects of Spam Accounts

The presence of spam and false accounts has negative effects on the quality of user experience and the general health of the platform. A plethora of negative effects of spam accounts have been documented through rigorous peer-reviewed research studies. Some studies documented illicit activities perpetrated by spam accounts via Twitter, such as scams and fraud, while others focused on deception and influence campaigns. I will document some of these studies next.

### 1.   Scams and Frauds

This form of abuse is rampant on Twitter. Spam accounts are often implicated in various forms of scams and frauds, most prominently in recent years (a) by means of cryptocurrency Ponzi schemes and/or pump-and-dump campaigns, (b) by promoting counterfeit and/or knock-off products, (c) by linking to websites selling illicit or unapproved drugs, (d) by advertising advance-fees scams, (e) by promoting phishing websites, (f) by linking to malware downloads, etc. It's worth noting that Twitter sanctions most of these behaviors but appears to fail to suspend all accounts that are involved in such behavior [Nizzoli et al., IEEE Access 2020; Phillips & Wilder, 2020; Nghiem et al., 2021; Eigelshoven, et al., 2021].

The authors of the ReadyToAbuse study [Mazza et al., 2022], were able to quantify the effects of at least one of the spam campaigns they discovered, that lead to stealing over 90 thousand dollars from unwitting Twitter users:

> *"Among our results is the identification of four campaigns based on coordinated inauthentic behavior (CIB), exposing attempts to influence the political debate and aggressive spamming aimed at deceiving users or advertising counterfeit products. We also uncovered that one of these campaigns managed to steal at least $90K from deceived users. Notably, the CIB instances adopted different strategies: amplify content through retweets; impersonate reputable accounts to infiltrate ongoing conversations; leverage the hype surrounding specific events to gain visibility; use third-party content to build a network and then spam own contents"*

In a recent study, Nizzoli and collaborators [Nizzoli et al., IEEE Access 2020], charted the landscape of cryptocurrency manipulations via social media. They investigated 5.7M Twitter users responsible for 16.8M crypto-related posts. These are some excerpts from the study (verbatim):

- "we collected a large dataset, composed of more than 50M messages published by almost 7M users on Twitter, Telegram and Discord, over three months.We performed bot detection on Twitter accounts sharing invite links to Telegram and Discord channels, and we discovered that more than 56% of them were bots or suspended accounts."

- "We uncover the pivotal role of Twitter bots in broadcasting invite links to deceptive Telegram and Discord channels [...] we observed that 93% of the invite links shared by Twitter bots point to Telegram pump-and-dump channels, shedding light on a little-known social bot activity."

- "we reported on 15 Twitter botnets that are responsible for the 75.4% of invite links to pump-and-dump channels"

**Figure 4** shows a botnet of spam accounts all promoting the same telegram crypto-scam account: it is evident that these accounts are part of the coordinated network; only a fraction of the accounts in this botnet was suspended by Twitter, whereas the majority of them (56.3%) remained active and engaged with the crypto spam campaign.

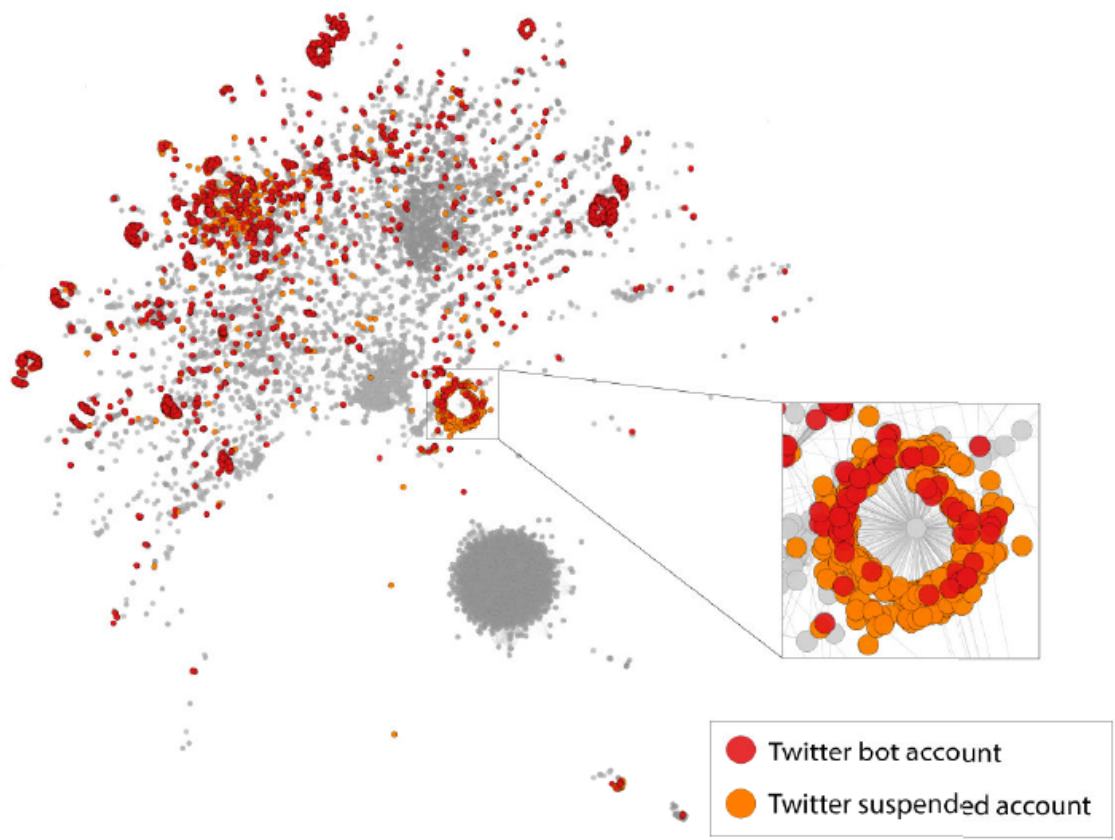

**Figure 4**. Invite link network highlighting deceptive Twitter accounts. A large portion of Twitter accounts has a deceptive nature (56.3%). The typical star structures frequently correspond to botnets promoting a single channel. We found 69 botnets with a size of at least 10 elements. From [Nizzoli et al., IEEE Access 2020]

## 2. Stock Market Manipulation

Reports of suspected influence of spam accounts on stock market transactions have been circulating for over a decade:

- Back in March 2011, Dan Mirvish, a Huffington Post contributor and media expert, linked the internet popularity of actor Anne Hathaway to spikes in the stock value of Warren Buffet's Berkshire-Hathaway, stating (verbatim) "When Anne Hathaway makes headlines, the stock for Warren Buffett's Berkshire-Hathaway goes up [...] It looks like all the automated, robotic stock trading programs are picking up the same chatter on the internet about "Hathaway" as the IMDb's StarMeter, and they're applying it to the stock market."[10] This was later named the Hathaway effect.

- Hwang and collaborators, who back in 2012 studied the Hathaway effect more broadly, postulated that spam bots on social media could exacerbate the effects of stock market manipulations [Hwang et al., Interactions 2012].

- The issue became such a concern that in November 2014, the SEC posted an alert on their bulletin titled "Investor Alert: Social Media and Investing" to bring awareness about social media stock manipulations and fraud schemes.[11]

Perhaps the most egregious example of the effect of Twitter spam on the stock market, and malicious use of spam accounts to pump and dump stocks, involves a little-known tech company called *Cynk*:

Seeking Alpha was first to report suspicious spam activity around the $CYNK stock (see **Figure 5**).[12] The NY Post later stated that "*Seeking Alpha found out early on about Cynk's otherwise inexplicable run that "no fewer than a dozen stock promoters" were hyping the stock in their tweets".[13]*

---

[10] The Hathaway Effect: How Anne Gives Warren Buffett a Rise - by the Huffington Post https://www.huffpost.com/entry/the-hathaway-effect-how-a_b_830041

[11]Updated Investor Alert: Social Media and Investing - Avoiding Fraud - by the Security and Exchange Commission (SEC) https://www.sec.gov/enforce/investor-alerts-and-bulletins/ia_socialmediafraud#.VQsIFBDF-mA

[12] Promoters Push Market Cap To $655 Million Despite $39 In Assets And No Revenue - by Seeking Alpha https://seekingalpha.com/article/2274553-cynk-technology-promoters-push-market-cap-to-655-million-despite-39-in-assets-and-no-revenue-100-percent-downside

[13] SEC sees price and reality not in 'Cynk' - by the New York Post https://nypost.com/2014/07/11/sec-sees-price-and-reality-not-in-cynk/

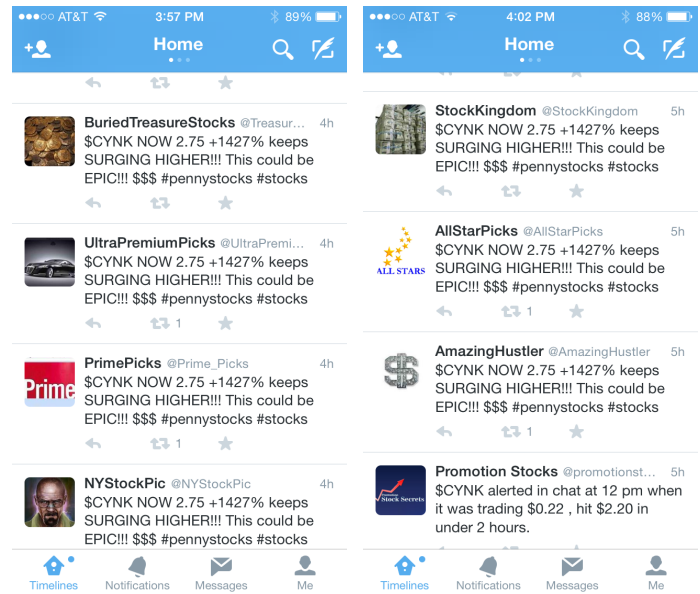

**Figure 5**: Spam accounts coordinately pumping the $CYNK stocks discourse on Twitter. From SeekingAlpha.

Forbes best summarized the stock market manipulation event as (verbatim):[14]

- "Last summer, a small, unknown social media company called Cynk Technology (CYNK) experienced the kind of runup that is the stuff of Wall Street legend. In less than a month, its stock price rose from six cents to nearly twenty-two dollars a share, an increase of more than 36,000 percent.".

- "The combination of bots and HFT algorithms in a market can amount to a conspiracy of virtual dunces, with bot programs pumping out fake "news" that will prompt an HFT algorithmic program to buy or sell, which sends a trading signal to other algorithmic programs, the whole thing snowballing until either until a completely worthless company is worth $6 billion or the entire market spirals into a crash, wiping out billions of dollars."

- "The SEC eventually suspended trading on CYNK, figuring it for a pump-and-dump, but by then the damage was done and the scammers had won. CYNK is now back to trading at a penny."

An independent analysis carried out by the analyst firm *Product Reviews* used Botometer version-1 (BotOrNot) stating that "*On a website called Bot or Not, we searched some of the Twitter accounts promoting this stock and it seems like most of these are robot accounts rather than humans.*".[15]

---

[14] The Dark Arts Of Social Media On Wall Street - by Forbes
https://www.forbes.com/sites/etrade/2015/05/26/the-dark-arts-of-social-media-on-wall-street/
[15] CYNK STOCK PRICE PUSHED HIGH BY TWITTER BOTS? - by Product Review:
https://www.product-reviews.net/cynk-stock-price-pushed-high-by-twitter-bots/

The stock market manipulation and related events were later investigated by the SEC and the perpetrator was charged with microcap fraud.[16]

In 2020, a study conducted by Tardelli and collaborators focused on characterizing financial disinformation campaigns on Twitter (verbatim) [Tardelli et al., HCII 2020]:

- "aim to shed light on this issue by investigating the activities of large social botnets in Twitter, involved in discussions about stocks traded in the main US financial markets. We show that the largest discussion spikes are in fact caused by mass-retweeting bots".

- "we focus on characterizing the activity of these financial bots, finding that they are involved in speculative campaigns aimed at promoting low-value stocks by exploiting the popularity of high-value ones".

- "These accounts appear as untrustworthy and quite simplistic bots, likely aiming to fool automatic trading algorithms rather than human investors."

- "Our findings pave the way for the development of accurate detection and filtering techniques for financial spam"

### 3. Political Disinformation

Efforts to distort political discourse via Twitter had been reported ever since the early 2010s [Ratkiewicz et al., WWW 2011; Metaxas & Mustafaraj, Science, 2012]. But on the onset of the 2016 U.S. presidential election, a rather new phenomenon was observed on Twitter, in concert with bots [Bessi & Ferrara; First Monday 2016], and hyper-partisan campaigns: the spread of false news and the coordination of disinformation campaigns [Allcott & Gentzkow, 2017; Marwick & Lewis, 2017; Mele, et al., 2017]. Disinformation is the deliberate spread of false or inaccurate information – to be distinguished from misinformation, where there is no planned agenda or intent to spread inaccuracies or falsehood. The adoption of automated accounts such as bots in the context of disinformation campaigns is particularly concerning because there is the potential to reach a critical mass large enough to dominate the public discourse and alter public opinion [Ferrara, et al., 2016; Woolley & Howard, 2016; Marwick & Lewis, 2017]; this could steer the public's attention away from facts and redirect it toward manufactured, planted information.

In August 2017, I published an analysis on disinformation and bot operations ahead of the 2017 French presidential election [Ferrara, First Monday 2017]. The study highlights (verbatim):

- *"We monitored the Twitter stream between 27 April and 7 May 2017 (Election Day), and collected a very large dataset containing nearly 17 million tweets related to the 2017 French presidential election. Within it, we identified the subset related to the MacronLeaks disinformation campaign."*

---

[16] SEC Charges Man With Microcap Fraud Involving Shares of Cynk Technology Corp. https://www.sec.gov/news/press-release/2015-157

- *"Out of 99,378 users involved in MacronLeaks, our model classified 18,324 of them as social bots, and the remainder of 81,054 as human users. The fraction of social bots amounts for about 18 percent of the total users involved in the campaign"*

- *"we uncovered that accounts used to support then-presidential candidate Trump before the 2016 U.S. election have been brought back from a limbo of inactivity (since November 2016) to join the MacronLeaks disinformation campaign. Such anomalous usage patterns point to the possible existence of a black market for reusable political-disinformation bots."*

- Here are some statistics regarding the performed activity and influence that the bot accounts accrued over the short 10 days before the election: (a) Bots posted on average 2.86 MacronLeaks-related tweets; (b) Bots obtained on average 1,382 followers ($\sigma$ = 22,282); (c) Bots friended on average 1,058 users ($\sigma$ = 12,190); (d) Bots favorited on average 228 tweets ($\sigma$ = 924); (e) Finally, bots have been listed on average on 7.42 lists ($\sigma$ = 90.3).

In 2020, Im and collaborators published the StillOutThere study [Im et al., 2020]. This study was aimed at assessing whether Russian trolls were still present and active on Twitter and the extent of their continued influence operations. The authors stated they (verbatim):

- *"develop machine learning models that predict whether a Twitter account is a Russian troll within a set of 170K control accounts; and,"*

- *"demonstrate that it is possible to use this model to find active accounts on Twitter still likely acting on behalf of the Russian state"*

### 4. Political Deception

Another form of influence exerted by spam accounts is political deception. According to [S. Woolley, 2016], [P. Howard et al., 2018], [S. Woolley & P. Howard, 2017], the behavior of political spam accounts can be categorized with respect to their intents:

- In political discussions, manufacturing consensus, to enhance the perception of popularity or influence of an entity (political actor, party, organization, etc.);
- Bolstering opinions and voices, to amplify the platform and audience that an entity will receive;
- Cementing polarization, by increasing the inflammatory or divisive nature of an issue or agenda;
- Increasing chaos and confusion, by posting inaccurate information, disinformation, and rumors;
- Algorithmic manipulation, to trick recommendation and ranking systems used by social media platforms, and give higher visibility to certain actors, viewpoints, or campaigns.

## 5. Conspiracy Amplification

The issue of *conspiracy amplification* has been studied extensively in recent years [Wang et al., 2022; Sharma et al., 2022]. In [Ferrara et al., First Monday 2020], we illustrated the role that bots played in amplifying content generated by alt-right or hyper-partisan news outlets during the 2020 US Presidential Election campaigns.

*Figure 6* shows that there is an unmistakable correlation between the fraction of bots engaged in the spread of a given link to a news site, and the likelihood of that news site being hyper-partisan or conspiratorial. The more conspiratorial or hyperpartisan a news site is, the more likely it is to see a substantial amount of bots spreading its content.

Hyper-partisan news outlets like One America News Network (OANN) and Infowars see the greatest proportion of their user base tweeting QAnon material. These two outlets also have the highest average bot scores. Left-leaning news outlets such as the New York Times and Washington Post have low bot scores and a proportion of QAnon users. The proportion of users using QAnon keywords is highly correlated with the average Botometer score (correlation coefficient: 0.947) across the spectrum of left, right, and neutral outlets.

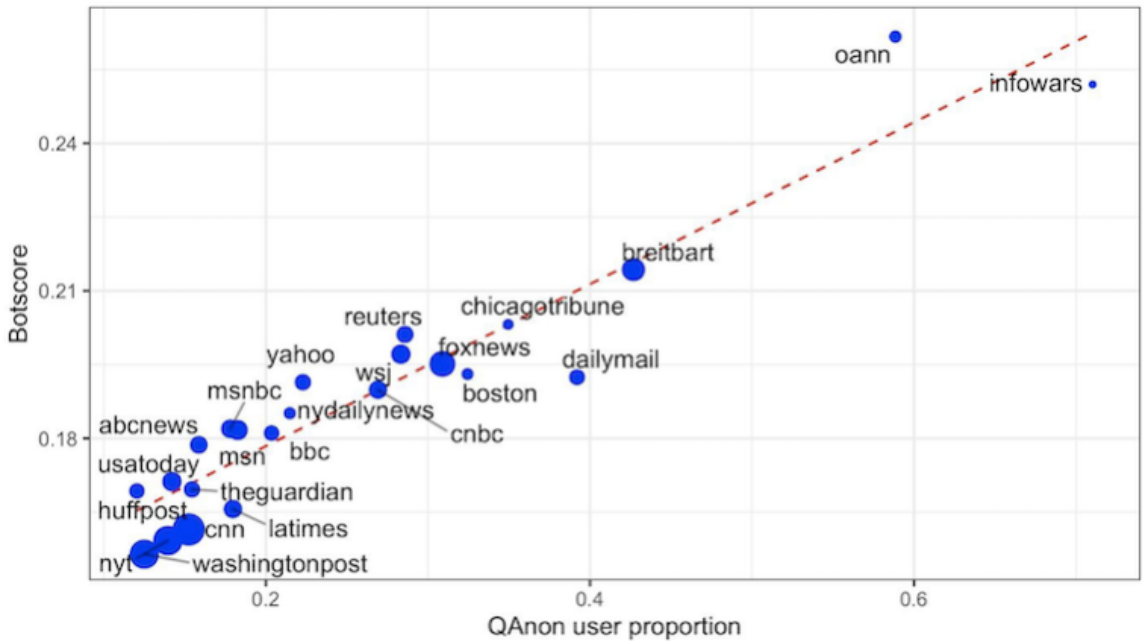

**Figure 6**: Proportion of users using QAnon hashtags and mean botscore for each news outlet, dot size indicates relative number of tweets. From [Ferrara et al.; First Monday 2020]

## 6. Coordinated Influence Campaigns

Spam and false accounts have been implicated in coordinated influence campaigns, as disclosed by Twitter as well as numerous research studies published in recent years. For example:

- Pacheco and collaborators investigated a coordinated influence campaign, later attributed to Russia, against the *Syria Civil Defense* (also known as the *White Helmets*) [Pacheco et al., WWW 2020] and reported that (verbatim):

  *"We unveil coordinated groups using automatic retweets and content duplication to promote narratives and/or accounts. The results also reveal distinct promoting strategies, ranging from the small groups sharing the exact same text repeatedly, to complex "news website factories" where dozens of accounts synchronously spread the same news from multiple sites."*

  Similar results were independently published by Wilson & Starbird [Wilson & Starbird, CSCW2 2021]: (verbatim)

  *"Our findings reveal a network of social media platforms from which content is produced, stored, and integrated into the Twitter conversation. We highlight specific activities that sustain the strategic narratives and attempt to influence the media agenda."*

- In 2021, Nizzoli and collaborators demonstrated the presence of coordinated influence campaigns in the context of the 2019 UK General Election [Nizzoli et al., ICWSM 2021]: (verbatim)

  *"We empirically demonstrate that coordination and automation are orthogonal concepts. Thus, our framework can complement long-studied techniques for detecting automation, manipulation, and inauthenticity."*

  The study demonstrates that coordinated behavior and automation (i.e., bot accounts) are two sides of the same coin (see **Figure 7**), and therefore different means of detection need to be developed and employed to guarantee the detection of orchestrated influence campaigns. Solely relying on bot detection techniques won't suffice.

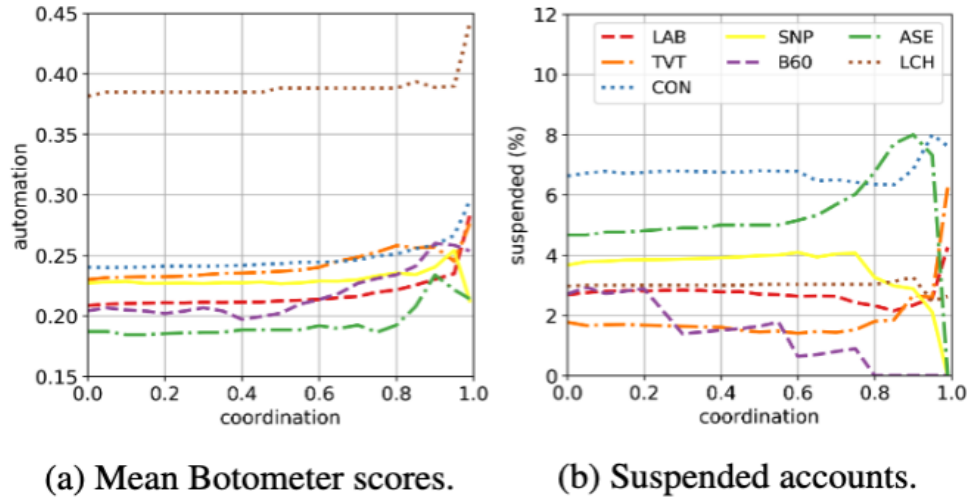

(a) Mean Botometer scores.    (b) Suspended accounts.

**Figure 7**: *"Correlation between coordination and use of automation, in terms of bot scores estimated by Botometer, and accounts suspended by Twitter. Both indicators of possible automation appear to be two orthogonal and largely uncorrelated with respect to coordination."* From [Nizzoli et al., ICWSM 2021].

- In [Sharma et al., KDD 2021], we demonstrated that the detection of coordinated accounts implicated in orchestrated influence campaigns is achievable by means of machine learning, obtaining accurate, scalable and reliable detection systems.

    *"We verified the effectiveness of the proposed method and training algorithm on real-world social network data collected from Twitter related to coordinated campaigns from Russia's Internet Research Agency targeting the 2016 U.S. Presidential Elections, and to identify coordinated campaigns related to the COVID-19 pandemic."*

    *"Leveraging the learned model, we find that the average influence between coordinated account pairs is the highest. On COVID-19, we found coordinated groups spreading anti-vaccination, anti-masks conspiracies that suggest the pandemic is a hoax and political scam."*

## 7. Public Health Misinformation

Researchers showed that spam and false accounts play a major role in the spread of public health misinformation. Numerous studies, both prior to the Covid19 pandemic and in the aftermath, demonstrated that rumors, unscientific or unsubstantiated claims, unsupported or unverified medical information, and public health misinformation thrive on Twitter in part, or largely due to the activity of spam accounts. For example:

- In 2014, in the midst of the "Ebola crisis", a study of mine portrayed the landscape of Twitter discussion about it [Ferrara; SIGWEB 2015]. We postulated that (verbatim):

    ***Figure 8*** *shows "the spread of concern and fear-rich content during the same interval of time (some example tweets are reported in **Table 5**). Positioning, size and colors*

*again represent the prominence of the accounts involved in the discussion and different Twitter topical communities. We can observe how panic and fear spread virally, reaching large audiences at great diffusion speed: the exposure to contents that leverage human's fears obfuscates our judgment and therefore our ability to discriminate between factual truths and exaggerated news. In turn, this fosters the spreading of misinformation, rumors, and unverified news, potentially creating panic in the population, and the generation of further, more negative content."*

We also proposed a strategy to tackle this issue: *"This self-reinforced mechanism can be interrupted by the implementation of ad-hoc intervention campaigns, designed to be effective on specific targets, based on their characteristics and susceptibility."*

| Time | Content |
|------|---------|
| 2014-10-17 13:33:57 | @FoxNews WHY WON'T YOU REPORT THE CRUISE SHIP IS BEING DENIED ENTRY INTO BELIZE? CRUCIAL FOR CLOSING OUR BORDERS! |
| 2014-10-17 13:39:44 | @FoxNews Those hc workers are totally wacky.Y today do they decide to tell some1 they handled specimens from Ebola pt, esp on ship¬ sick? |
| 2014-10-17 13:40:38 | @NBCNews Many countries are refusing entry from flights from Ebola nations. Why aren't you asking the WH about their crazy policy? |
| 2014-10-17 13:56:20 | @CNN Why two doctors treated for EBOLA & INFECTED no one else including CARE GIVERS but it's not the same case for TX? Who screwed up? |
| 2014-10-17 13:59:20 | @FoxNews Are they paying these potential Ebola victims money to get on ships and planes? I am beginning to really wonder |
| 2014-10-17 14:09:03 | OMG! He cld b infected "@NBCNews: Who is man wearing plain clothes during an #Ebola patient's transfer?" via @NBCDFW |

**Table 5**. Examples of concerned fear-rich tweets spreading during the Ebola emergency of Oct. 17th, 2014.

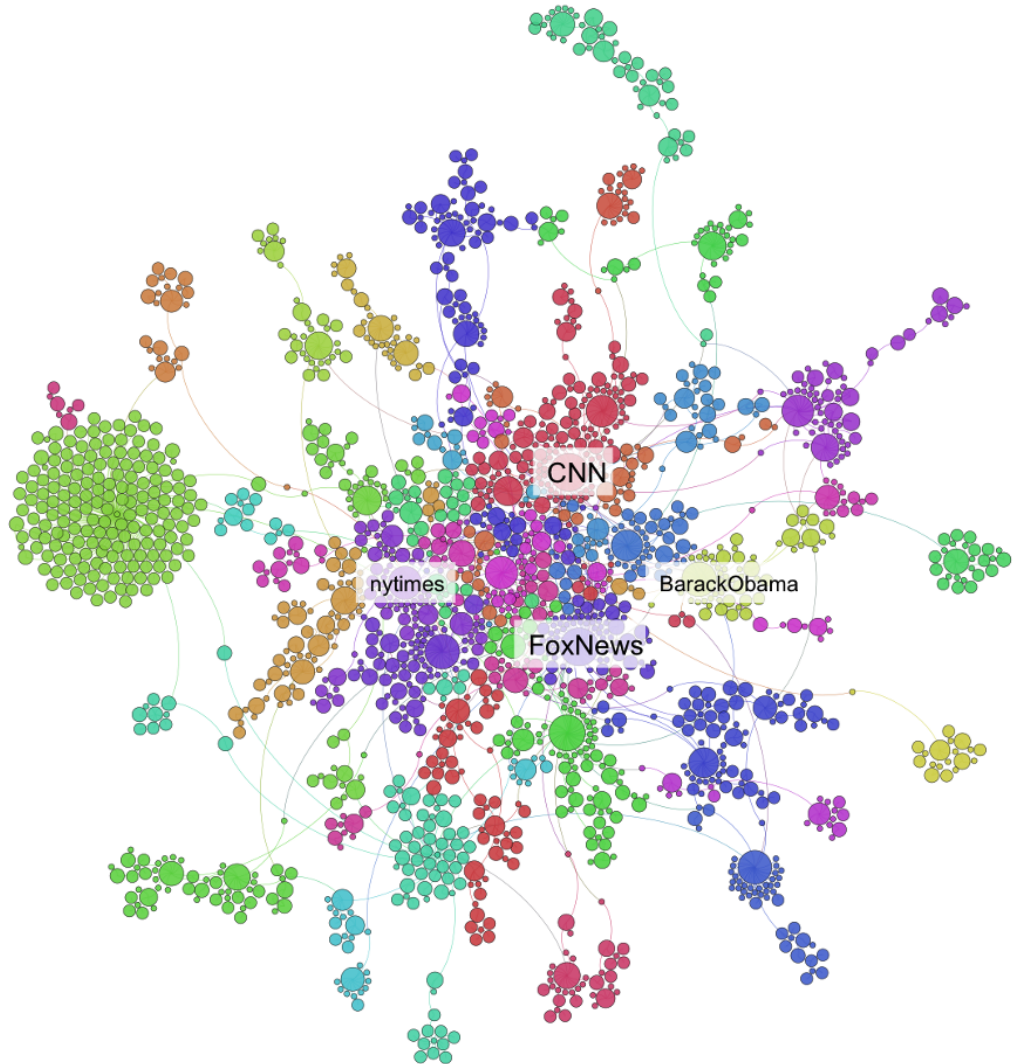

**Figure 8**: The spreading of fear-rich content across different Twitter communities (retweets and mentions). From [Ferrara, SIGWEB 2015].

- In 2018, Broniatowski and colleagues set (verbatim) *"To understand how Twitter bots and trolls ("bots") promote online health content."* [Broniatowski et al., AJPH 2018].

  They (verbatim) *"compared bots' to average users' rates of vaccine-relevant messages, which we collected online from July 2014 through September 2017. We estimated the likelihood that users were bots, comparing proportions of polarized and antivaccine tweets across user types. We conducted a content analysis of a Twitter hashtag associated with Russian troll activity."*

  They discovered statistically significant evidence that *"Compared with average users, Russian trolls, sophisticated bots, and "content polluters" tweeted about vaccination at higher rates. Whereas content polluters posted more antivaccine content, Russian trolls amplified both sides. Unidentifiable accounts were more polarized and antivaccine. Analysis of the Russian troll hashtag showed that its messages were more political and divisive."*

> *"Malicious online behavior varies by account type. Russian trolls and sophisticated bots promote both pro- and antivaccination narratives. This behavior is consistent with a strategy of promoting political discord. Bots and trolls frequently retweet or modify content from human users."*

This study demonstrated public health misinformation campaigns (later attributed to Russia) that leveraged spam accounts, bot accounts, sophisticated bots, and trolls, disproportionately engaged with the antivaccine discourse on Twitter (cf., **Figure 9**). Second, such bots and trolls aimed to sow discord by amplifying both pro- and anti-vax arguments causing confusion and disagreement among other users.

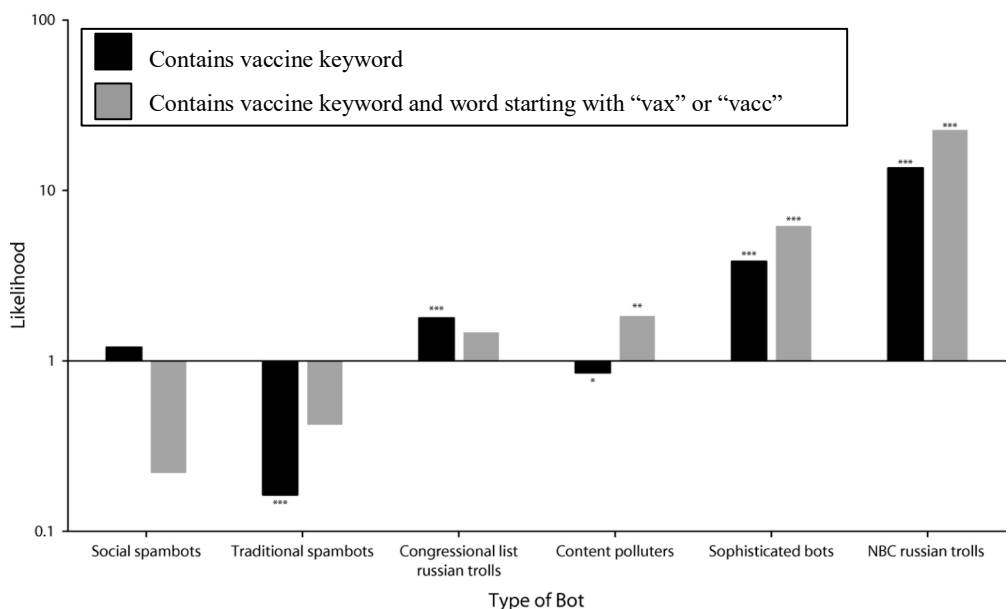

**Figure 9**: Bots' Likelihood of Tweeting About Vaccines Compared with Average Twitter Users: July 14, 2014–September 26, 2017. From [Broniatowski et al.; AJPH 2018].

- Since early 2020, with the Covid19 pandemic, a plethora of studies (simply too many to discuss in this context) investigated spam, false, and bot accounts, focusing on their role and effects contributing to the spread of public health misinformation related to Covid19. The evidence is overwhelmingly in support of the theory that spam and false accounts had nefarious effects on the spread of misinformation about Covid19. A recent survey by the *World Health Organization* (WHO) tried to summarize some findings [Gabarron et al., 2021]. Since the, numerous other studies have been published that provide further evidence of Covid-related public health misinformation on Twitter [Chen et al., 2021; Muric et al., 2021; Rao et al., 2021; Jiang et al., 2021; Chen et al., 2022].

## 8. Crisis Response Disruption

Researchers showed that Twitter acts as a responsive social sensor in times of crises, reflecting the information-seeking efforts of people involved in different kinds of emergencies or natural disasters, such as earthquakes, tsunamis, fires, hurricanes, floods, etc. [Sasaki et al., WWW 2010; Yates et al., IJIM 2011; Yin et al., IJCAI 2015].

However, what happens if the conversation in the context of an emergency is polluted by spam? Various studies have investigated the effect of spam and false accounts in the context of crises.

- Gupta and collaborators, back in 2013, were among the first to study the effect of false rumors spreading after the Boston Marathon bombing of 2012 [Gupta et al., APWG 2013]. They reported (verbatim):

  *"We analyzed one such media i.e. Twitter, for content generated during the event of Boston Marathon Blasts, that occurred on April, 15th, 2013. A lot of fake content and malicious profiles originated on Twitter network during this event."*

  *"Our results showed that 29% of the most viral content on Twitter, during the Boston crisis were rumors and fake content; while 51% was generic opinions and comments; and rest was true information."*

  **Figure 10** illustrates the coordinated strategies suspected accounts employ to spread rumors or spam in the context of crises.

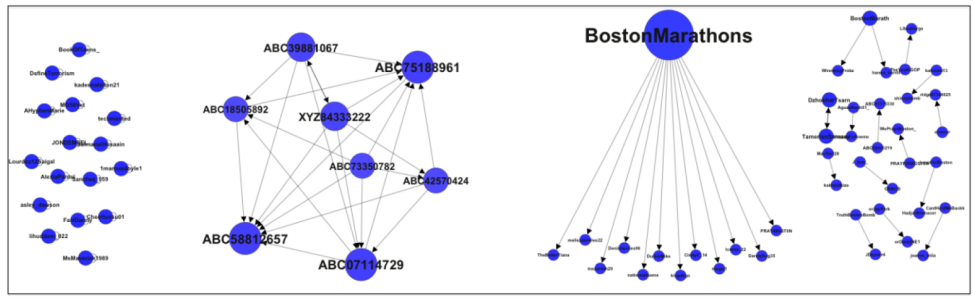

**Figure 10**: *"Network of suspended accounts (retweets / replies / mentions) created during the Boston blasts. We see four different forms of interactions amongst the suspended profiles (left to right): Single Links, Closed Community, Star Topology and Self Loops."* From [Gupta et al.; APWG 2013]

- In 2015, Zhao and collaborators published a study of rumors on Twitter [Zhao et al., WWW 2015]. They developed a technique to detect clusters of Twitter posts whose topic is a disputed factual claim. They found that (verbatim):

  *"On a typical day of Twitter, about a third of the top 50 clusters were judged to be rumors, a high enough precision that human analysts might be willing to sift through them."*

  *"One week after the Boston bombing, the official Twitter account of the Associated Press (AP) was hacked. The hacked account sent out a tweet about two explosions in the White House and the President being injured"* and *"within 60 seconds after the hacked AP account sent out the rumor about explosions in the White House, there were already multiple users enquiring about the truth of the rumor (**Figure 11**). Table shows some examples of these enquiry tweets."*

  **Remarks**: this study is important because first, it quantifies in ⅓ the amount of rumor clusters among Twitter discussions revolving around a crisis, an alarmingly large amount of spam or false content. Second, it shows evidence that post-hoc mitigation strategies based on suspension are ineffective since rumors spread virally within minutes.

| |
|---|
| Oh my god is this real? RT @AP: Breaking: Two Explosions in the White House and Barack Obama is injured |
| Is this true? Or hacked account? RT @AP Breaking: Two Explosions in the White House and Barack Obama is injured |
| Is this real or hacked? RT @AP: Breaking: Two Explosions in the White House and Barack Obama is injured |
| How does this happen? #hackers RT @user: RT @AP: Breaking: Two Explosions in the White House and Barack Obama is injured |
| Is this legit? RT @AP Breaking: Two Explosions in the White House and Barack Obama is injured |

**Figure 11**: False tweets about an alleged explosion in the White House, a hack later attributed to the Syrian Electronic Army. Within seconds, the rumor went viral. From [Zhao et al., WWW 2015].

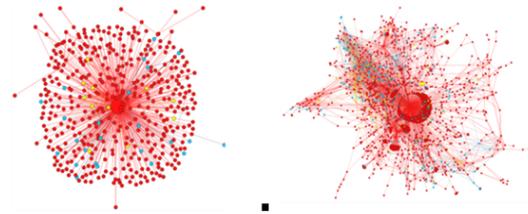

(a) 60 seconds after the hacked twitter account sent out the White House rumor there were already sufficient enquiry tweets (blue nodes).

(b) Two seconds after the first denial from an AP employee and two minutes before the official denial from AP, the rumor had already gone viral.

## 9.    Extremism, Radical Propaganda, and Recruitment

In 2015, Berger and Morgan documented the myriads of efforts and associated strategies that ISIS and other extremist groups adopted on Twitter [Berger & Morgan; 2015]. The authors observed that:

> "ISIS uses several practices designed to amplify its apparent support on Twitter, including "bots" (computer software that creates activity on a social media account in the absence of a human user) and spam (purchased tweets promoting ISIS content). We were able to compensate for this, although not perfectly."

> "In the overall Census Dataset, around 400 nonclient apps were detected to be in use among more than 6,000 accounts. Within the 5.4 million Demographics Dataset tweets analyzed, hundreds of additional bots and apps were also detected operating at lower prevalences, enough to suggest that perhaps 20 percent or more of all tweets in the set were created using bots or apps."

## Conclusions and Recommendations

**Why is Twitter the emblematic use case to study social media manipulation and abuse?** In this survey, I aimed to provide a (non-comprehensive) summary of the literature that appeared over the last decade or so that tackles the problem of abuse and manipulation of social media, with a focus on spam and false accounts on Twitter. I argued that dealing with this problem is of paramount importance and that Twitter represents the most emblematic use case as it suffered from many forms of abuse that had major societal implications.

In reviewing the research discussed above, I highlighted how Twitter has suffered from the most prototypical forms of abuse. Whether it was the study of the spread of misinformation and conspiracies, the detection of financial scams and frauds aimed to depauperate its users, the unveiling of state-sponsored influence campaigns, or toxic and hateful speech, Twitter has been the go-to choice for researchers and by far the most studied platform in the current social media landscape. Although naturally, Twitter is not the sole platform to suffer from these problems, I argued that it can serve as a testbed to understand and combat these problems, which can lead to developing tools and techniques that can then be applied to other mainstream platforms.

In the following, I will provide a set of recommendations that will hopefully pave the way for an agenda of future research, cognizant policy and decision-making, and an informed public.

## Technological Solutions

**Fostering data access**. Arguably, at least in part due to the ease of accessing its data, Twitter has been the subject of much academic (and beyond) scrutiny and studies [Williams et al., 2013]. Albeit this might on the surface appear to be detrimental to the public-facing image of the platform, I argue that instead Twitter has been at the forefront in terms of practices fostering data access to the research community at large. The ability to access in a systematic way, both in real-time and via search, massive droves of data via the Twitter APIs has enabled thousands of studies, which in turn allowed us to garner a better understanding not only of the problems that Twitter has been suffering from, but also of their implications, and some possible solutions. In this regard, open data practices similar to what Twitter has been doing for years should be cherished, and I for one would welcome more platforms embracing the same open data policies.

**The need for transparent algorithmic practices**. Data access is a necessary but not sufficient condition to enable the study and understanding of social media manipulation and abuse. Platforms employ increasingly complex algorithms to curate content, rank newsfeeds, recommend interactions, offer personalized information, serve targeted ads, etc. The need for transparency grows with the complexity of solutions, in particular for transparent practices enabling algorithmic auditing [Venkatadri et al., 2019; Raji et al., 2020]. Twitter, much like many other mainstream platforms [Ali et a., 2019; Ali et a., 2021], suffers from various forms of algorithmic biases [Bartley et al., 2021]. These issues can possibly be exacerbated by the fact that platforms continually operate as  massive-scale experiments, where A/B testing of different functionalities leads to different subsets of users being served slightly different versions of the platform, which can lead to new, unforeseen problems only affecting certain population subgroups [Hunter & Evans, 2016]. These issues are incredibly hard to study from an outsider's perspective, without a direct pathway to transparently access and audit the algorithms and the experiments carried out by the platforms.

**A place for artificial intelligence and machine learning.** The Artificial Intelligence / Machine Learning (AI/ML) research community has made giant leaps in the development of accurate and scalable techniques to identify different forms of platform abuse, including false and spam accounts. Hundreds, maybe thousands of peer-reviewed papers are published each year that propose a variety of approaches to leverage advances in AI/ML to tackle complex problems the like as detecting spam, compromised accounts, state-sponsored troll farms, bot and automated accounts, influence campaigns, extremist propaganda, cryptocurrency manipulation, and so on. Many of these techniques are also tested and evaluated by third-party independent scrutiny, and their limitations and strengths are well investigated. Some of these approaches are provided as services (e.g., as publicly accessible APIs), and some are continuously maintained and even improved over time. These are some valid rationales to consider AI/ML solutions to tackle such complex problems at the scale some of these massive platforms like Twitter operate.

**An online "Real ID" and other identity verification systems**. Although AI/ML might provide significant help in mitigating some issues, arguably some challenges in this space are harder than others. For example, for platforms that do not impose a strict enforcement (a la Facebook) on an account's identity matching an individual's identity, guaranteeing the integrity and authenticity of user accounts in other ways is paramount. This is in part because research has systematically found that, hiding behind the anonymity veil, users tend to misbehave or act in anti-social manners significantly more than when their identity is known. Anonymity comes with a high price, to quote Davenport who two decades ago stated that "*by allowing anonymous communication we actually risk an incremental breakdown of the fabric of our society*" [Davenport, 2002]. I recently postulated that to certify user identity and information veracity we might rely on blockchain technologies [Ferrara, 2019]. Since then, some approaches inspired by the same principles, such as *humanID*,[17] have been proposed to verify and certify users' identity while keeping them distinctly separated from publicly-visible account information of a platform.

## Social and Policy Solutions

**The urge for responsible policymaking.** Technology-based solutions might help make a dent in the problems above but are bound to be insufficient, and history shows they are likely to fail if not backed up by sensible policymaking. Twitter has been under public scrutiny and subject to multiple kinds of investigations, as did other mainstream platforms like Youtube and Facebook. But the expectation that social media companies might be able to implement magic bullet solutions to complex problems like abuse and manipulation is not bound in reality. Policymakers must enact proposals that are sensible and rooted in scientific consensus, and ascertain what is feasible to expect in terms of technological solutions, and what instead requires new legislature and new policies. This is where corporations, researchers, and law and policymakers need to cooperate for the greater good. One successful example of such convergence revolves around the problem of regulating social media bots: the US Senate Investigation into the 2016 US Presidential Election helped expose the now well-known foreign-sponsored efforts led by Russia, which revealed a need for legislating bots. The California B.O.T. Act ("Bolstering Online Transparency") and the Senate Bill "Bot Disclosure and Accountability Act of 2019", have been since formulated, and the former became operational since July 2019, to regulate what activities automated accounts can and cannot engage with in the context of online political discussion.

**Academic and corporate social responsibility.** Governments, however, do not have the technical and scientific capabilities and knowhow to face these big challenges with top down approaches. A need arises for socially responsible industry standards to tackle these problems. Without close collaborations between researchers in academia, industry, and third-party practitioners, the hope is faint to come full circle and propose socio-technical solutions that can work in combating the problem of social media spam and abuse. A major obstacle to success is the fact that some abuse practices are not necessarily technologically detectable (e.g., requests to retweet, trading followers, directed-message persuasion, etc.): an equitable algorithmic and socially responsible information ecosystem cannot come to fruition without a set of shared rules aimed at normalizing what behaviors are tolerated and what represent abuse, rather than *ad hoc* Terms of Service rules that change from platform to platform.

---

[17]Foundation for a Human Internet: http://human-internet.org/

**The role of research sponsors and funders.** Much responsibility has been put onto the academic and industrial research endeavor to come up with socio-technical solutions to these problems. Nevertheless, I argue that we deal with pressing societal issues and social media spam and abuse can lead to tangible, nefarious impacts for democracy and society at large. Hence, our research must be at the top of public funding agendas, in stark contrast with the limited number of projects and programs that have been devoted to sponsoring research into this problem. It is in the public interest that social media abuse and manipulation is mitigated, and not solely the responsibility of platforms to ensure so. Therefore, supporting research into socio-technical solutions with public funds is warranted now more than ever.

**Our future might also depend upon an educated public.** Whether major problems like misinformation can ever be solved with the sole use of socio-technical solutions has been subject of much academic debate. Hence, there is a role for literacy toward an educated internet-based hyper-connected society. But it would be wishful thinking to expect that literacy would suffice to defeat problems like social media manipulation. For example, thanks to the leaps in Artificial Intelligence, it is nowaday nearly impossible for a non-trained human to distinguish between organic and machine-generated content (e.g., a short, tweet-like text, or a profile picture): AI techniques based on *large generative models* for text (e.g., GPT3 and its successors) and multimedia (DALL-E 2, Midjourney, Stable Diffusion, etc.) can trick the human judgment, so even if a literate or tech-savvy user is exposed to such manipulated or AI-generated content, there is a limit to our cognitive and judgemental abilities to ascertain whether what we see is true or machine-generated, or whether what we read is factual or fabricated.

Literacy will prepare the future generation to question everything online, but the full trifecta of socio-technical solutions, sensible policy making, and an informed public will be necessary to truly uproot the problems of social media spam and abuse, and their possibly devastating effects.

## About the Author

Emilio Ferrara is Associate Professor at the University of Southern California, Research Team Leader at the USC Information Sciences Institute, and Principal Investigator at the USC/ISI Machine Intelligence and Data Science (MINDS). Ferrara uses AI for modeling and predicting human behavior in techno-social systems and has published 200+ papers on these topics. Ferrara served as the trial expert on the issue of spam and false accounts for the Delawere Court of Chancery in the Twitter vs. Musk litigation.

## Acknowledgements

The author is indebted with Professor Tim Weninger (University of Notre Dame), and Professors Patrick Warren and Darren Linvill (Clemson University) for their invaluable feedback.